\documentstyle[epsf]{mn}
\newif\ifAMStwofonts

\ifoldfss
  \ifCUPmtlplainloaded \else
    \NewTextAlphabet{textbfit} {cmbxti10} {}
    \NewTextAlphabet{textbfss} {cmssbx10} {}
    \NewMathAlphabet{mathbfit} {cmbxti10} {} 
    \NewMathAlphabet{mathbfss} {cmssbx10} {} 
  \fi
  \ifAMStwofonts
    \ifCUPmtlplainloaded \else
      \NewSymbolFont{upmath} {eurm10}
      \NewSymbolFont{AMSa} {msam10}
      \NewMathSymbol{\upi}     {0}{upmath}{19}
      \NewMathSymbol{\umu}     {0}{upmath}{16}
      \NewMathSymbol{\upartial}{0}{upmath}{40}
      \NewMathSymbol{\leqslant}{3}{AMSa}{36}
      \NewMathSymbol{\geqslant}{3}{AMSa}{3E}

       \let\le=\leqslant
       \let\ge=\geqslant
    \fi
  \fi
\fi 

\ifnfssone
  \newmathalphabet{\mathit}
  \addtoversion{normal}{\mathit}{cmr}{m}{it}
  \addtoversion{bold}{\mathit}{cmr}{bx}{it}
  \newmathalphabet{\mathbfit} 
  \addtoversion{normal}{\mathbfit}{cmr}{bx}{it}
  \addtoversion{bold}{\mathbfit}{cmr}{bx}{it}
  \newmathalphabet{\mathbfss} 
  \addtoversion{normal}{\mathbfss}{cmss}{bx}{n}
  \addtoversion{bold}{\mathbfss}{cmss}{bx}{n}
  \ifAMStwofonts
    \ifCUPmtlplainloaded \else
      \UseAMStwoboldmath
      \makeatletter
      \new@mathgroup\upmath@group
      \define@mathgroup\mv@normal\upmath@group{eur}{m}{n}
      \define@mathgroup\mv@bold\upmath@group{eur}{b}{n}
      \edef\UPM{\hexnumber\upmath@group}
      \new@mathgroup\amsa@group
      \define@mathgroup\mv@normal\amsa@group{msa}{m}{n}
      \define@mathgroup\mv@bold\amsa@group{msa}{m}{n}
      \edef\AMSa{\hexnumber\amsa@group}
      \makeatother
      \mathchardef\upi="0\UPM19
      \mathchardef\umu="0\UPM16
      \mathchardef\upartial="0\UPM40
      \mathchardef\leqslant="3\AMSa36
      \mathchardef\geqslant="3\AMSa3E

       \let\le=\leqslant
       \let\ge=\geqslant
    \fi
  \fi
\fi 

\ifnfsstwo
  \DeclareMathAlphabet{\mathbfit}{OT1}{cmr}{bx}{it}
  \SetMathAlphabet\mathbfit{bold}{OT1}{cmr}{bx}{it}
  \DeclareMathAlphabet{\mathbfss}{OT1}{cmss}{bx}{n}
  \SetMathAlphabet\mathbfss{bold}{OT1}{cmss}{bx}{n}
  \ifAMStwofonts
    \ifCUPmtlplainloaded \else
      \DeclareSymbolFont{UPM}{U}{eur}{m}{n}
      \SetSymbolFont{UPM}{bold}{U}{eur}{b}{n}
      \DeclareSymbolFont{AMSa}{U}{msa}{m}{n}
      \DeclareMathSymbol{\upi}{0}{UPM}{"19}
      \DeclareMathSymbol{\umu}{0}{UPM}{"16}
      \DeclareMathSymbol{\upartial}{0}{UPM}{"40}
      \DeclareMathSymbol{\leqslant}{3}{AMSa}{"36}
      \DeclareMathSymbol{\geqslant}{3}{AMSa}{"3E}

       \let\le=\leqslant
       \let\ge=\geqslant
    \fi
  \fi
\fi 

\ifCUPmtlplainloaded \else
  \ifAMStwofonts \else 
    \def\upi{\pi}
    \def\umu{\mu}
    \def\upartial{\partial}
  \fi
\fi

\title[KPD 0422+5421]{Confirmation of Eclipses in KPD 0422+5421, 
A Binary Containing a
White Dwarf and a Subdwarf B Star}
\author[Orosz \& Wade]{Jerome A. 
Orosz\thanks{Visiting Astronomer at Kitt Peak National 
Observatory (KPNO), which is operated by AURA, Inc., under a cooperative 
agreement 
with the National Science Foundation.} 
and Richard A. Wade \\
Department of Astronomy \& Astrophysics, The
Pennsylvania State University, 525 Davey Laboratory, \\
University Park, PA 16802-6305, USA}
\date{\today}
\pagerange{\pageref{firstpage}--\pageref{lastpage}}
\pubyear{1999}

\begin{document}

\maketitle

\label{firstpage}

\begin{abstract}
We report additional photometric CCD observations of KPD 0422+5421, a
binary with an orbital period of 2.16 hours which contains a subdwarf
B star (sdB) and a white dwarf.  There are two main results of this
work.  First, the light curve of KPD 0422+5421 contains two distinct
periodic signals, the 2.16 hour ellipsoidal modulation discovered by
Koen, Orosz, \& Wade (1998) and an additional modulation at 7.8
hours.  This 7.8 hour modulation is clearly not sinusoidal: the rise
time is about 0.25 in phase, whereas the decay time is 0.75 in phase.
Its amplitude is roughly half of the amplitude of the ellipsoidal
modulation.  Second, after the 7.8 hour modulation is removed, the
light curve folded on the orbital period clearly shows the signature
of the transit of the white dwarf across the face of the sdB star {\em
and} the signature of the occultation of the white dwarf by the sdB
star.   We used the Wilson-Devinney code to model the light
curve to obtain the inclination, the mass ratio, and the $\Omega$
potentials, and a Monte Carlo code to compute confidence limits
on interesting system parameters.  We find component masses of
$M_{\rm sdB}=0.36^{+0.37}_{-0.16}\,M_{\odot}$ and
$M_{\rm WD}=0.47^{+0.18}_{-0.16}\,M_{\odot}$
($M_{\rm total}=0.86^{+0.52}_{-0.35}\,M_{\odot}$, 68 per cent confidence
limits).  If we impose an additional constraint and require the
computed mass and radius of the white dwarf to be consistent with a
theoretical mass-radius relation, we find 
$M_{\rm sdB}=0.511^{+0.047}_{-0.050}\,M_{\odot}$ and
$M_{\rm WD}=0.526^{+0.033}_{-0.030}\,M_{\odot}$ 
(68 per cent confidence limits).
In this case the total mass of the system is less than $1.4\,M_{\odot}$
at the 99.99 per cent confidence level.
We briefly
discuss possible interpretations of the
7.8 hour modulation and the importance of KPD 0422+5421 as a member of
a rare class of evolved binaries.
\end{abstract}

\section{Introduction}
KPD 0422+5421 was first identified in the Kitt Peak Downes survey of
blue stars at mid to low galactic latitudes (Downes 1986).  It was
classified as a hot, hydrogen-rich subdwarf B (sdB) star.  The binary
nature of this star was discovered by Koen, Orosz, \& Wade (1998,
hereafter KOW) who determined an orbital period of $P=0.0901795 \pm
(3\times 10^{-7})$ days (2.16 hours) using spectroscopic and
photometric observations.  KOW showed that the companion star of the
sdB is a white dwarf, and determined masses of $M_{\rm sdB}=0.72\pm
0.26\,M_{\odot}$ and $M_{\rm WD}=0.62\pm 0.18\,M_{\odot}$ for the sdB
star and white dwarf, respectively.  KPD 0422+5421 has one of the
shortest known orbital periods for a detached binary, and is one of
only a few known binaries which contain an sdB star paired with a
white dwarf.  This binary has had an interesting past, given that it
contains two evolved stars in a close orbit, and will have an
interesting future, given the relatively short time-scale for merger
due to the loss of orbital angular momentum via gravitational wave
radiation ($\approx 1.5\times 10^8$ yr), which is comparable to the
core He burning lifetime of an sdB star (Dorman, Rood, \& O'Connell
1993).

The $U$ and $B$ light curves of KPD 0422+5421 presented in KOW are
ellipsoidal, with amplitudes of $\approx 0.02$ mag.  Ellipsoidal model
fits to the light curves using the Wilson-Devinney (1971, hereafter
W-D) code indicated a high inclination ($78^{\circ}$) which should
produce an eclipsing geometry.  The signature of the white dwarf
transit across the face of the sdB star
was predicted to be $\approx 0.006$ mag deep, lasting $\approx
27^{\circ}$ in phase ($\approx 10$ minutes of time).  The presence
of eclipses would be important since one could in principle then determine
the geometry of the system reasonably well.  However, the $U$ and $B$
light curves of KOW, which were obtained using high-speed
photomultipliers, were sufficiently noisy that this subtle signature
could not be clearly seen.  We therefore obtained additional
photometric observations using a CCD with the goal of obtaining a
precise eclipse profile.  We discuss below our observations, data
reduction and analysis, and our results.

\section{Observations and Data Reduction}

We observed KPD 0422+5421 December 27-30, 1998 (all dates are
UT) using the Kitt
Peak 2.1 metre telescope and the T1KA 
$1024\times 1024$ CCD.  The CCD was binned
$2\times 2$ on chip, yielding a readout time of $\approx 11$ seconds
and a scale of 0\farcs 6 per pixel.  Nearly all of the observations were
made with a standard Kron
$R$ filter (we used the
KPNO ``Harris set'' {\hbox{$U\!BV\!R$}}$I$ filters).  
The exposure times ranged from 20 to 40 seconds.
Care was taken to place the image of KPD 0422+5421 on the same region
of the detector in order to minimize possible systematic errors
introduced by flat-fielding errors.  The observations on the
first night were mostly taken through thin cirrus, while the
other three nights were clear.  The seeing was typically 1\farcs 0
to 1\farcs 2, although on some occasions the seeing was as low
as 0\farcs 75.  We obtained a total of 2676 $R$-band images of
KPD 0422+5421 covering 9.17 hours,
8.83 hours, 8.75 hours, and 9.83 hours on December 27 through 
December 30, respectively.  On December 28 we also obtained
images of a total of 38 standard stars in four different fields 
(RU 149, RU 152, PG 0918+029, PG 1047+003) from
the list of Landolt (1992) in the $U$, $B$, $V$, $R$, and $I$ filters
(airmass range 1.19 to 1.97),
as well as three sets of images of KPD 0422+5421 in the same filters
(airmasses of 1.44, 1.46, and 1.63).

Standard IRAF\footnote{IRAF is distributed by the
National Optical Astronomy Observatories.} 
tasks were used to remove the
electronic bias and to perform the flat-fielding corrections.
The programs {\sc DAOPHOT IIe}, {\sc ALLSTAR} and
{\sc DAOMASTER} 
(Stetson 1987; 
Stetson, Davis, \& Crabtree 1991; 
Stetson 1992a,b) were used
to compute the photometric time series of KPD 0422+5421 and all of the
field stars.  
The {\sc DAOPHOT} and {\sc ALLSTAR} codes were used because
KPD 0422+5421 has a faint neighbour star $\approx 2\farcs 3$ to the south
($\Delta R=5.1$) and because many of the bright field stars
also have near neighbours.
We used an iterative procedure to arrive at the final light curve.
First, an automated {\sc DAOPHOT/ALLSTAR}
script was used 
to determine the positions and
instrumental magnitudes of all the stars for each frame.
Then using the stellar positions determined above,
the {\sc DAOMASTER} code was used to compute the 
mean pixel offsets for each image with respect to the first one.
The pixel shifts were rounded to the nearest integer 
(to avoid interpolation in the shifting routine) and 
the images were then aligned so that they all had a common coordinate system.
A deep ``master'' image was made by combining 101 images where the seeing was
better than $\approx 0\farcs 9$, and
{\sc DAOPHOT} and {\sc ALLSTAR} 
were used to reduce this master image interactively.
In the end we selected nineteen relatively bright and isolated
field stars which gave the best fits
to the empirical point spread function (PSF).   The final ``master photometry
list'' for this deep image
had 838 stars, compared to the $\approx 300$ stars typically found
by the {\em daofind} routine on any given single image.  Next,
the aligned images were reduced using another automated {\sc DAOPHOT/ALLSTAR}
script.  
In this case, 
the stars from the master photometry list were used in
place of the initial stellar list generated by 
{\em daofind} for each image and
the same 19 stars were used to fit the PSFs for each image.  
We found that the script using the larger master photometry list and
common PSF stars gave much better results than the simple script used for
the first iteration.  Finally, the 
{\sc DAOMASTER} code was used to robustly
compute the mean magnitude offsets for each image and correct the magnitude
system of each image to the magnitude system of the first image by
adding a constant.

\begin{figure*}
\vspace{10.7cm}
\includegraphics{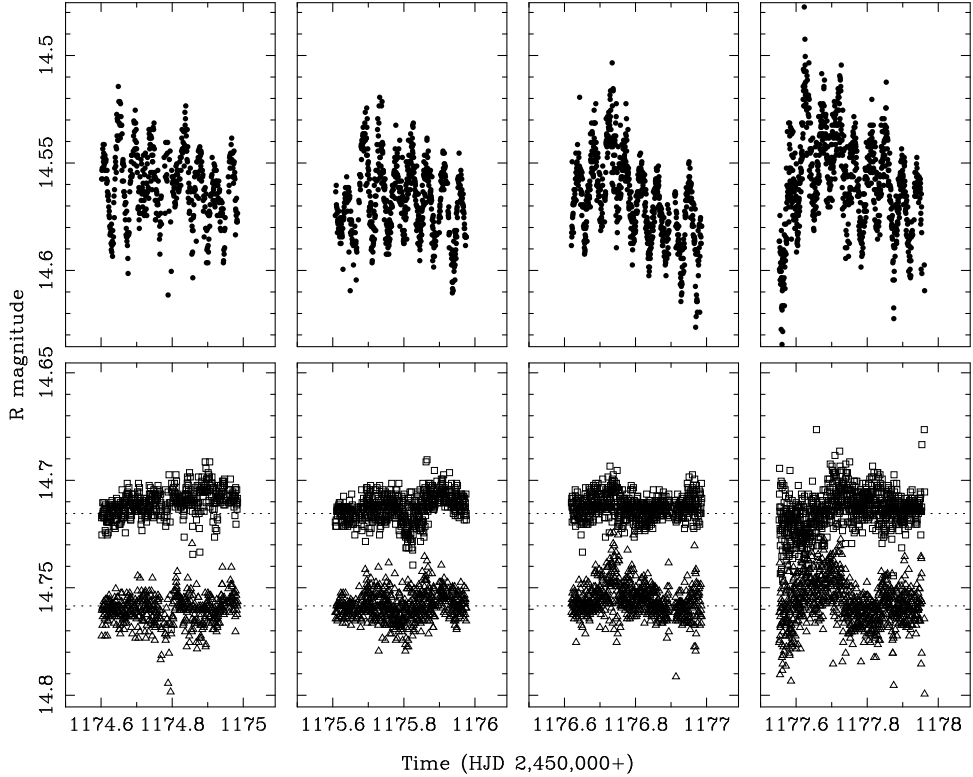} 
\caption{Top panels: the time series photometry of KPD 0422+5421
($R$-band).  In addition to the ellipsoidal modulation at 2.16 hours,
a signal with a much longer period is evident.
Bottom panels: the
light curves of two nearby comparison stars, plotted using the same
range in the $y$-axis as the top panels.
}
\label{fig1}
\end{figure*}

The observations of
the Landolt standard stars were used to calibrate
the {\sc DAOMASTER} instrumental magnitudes to the
standard system.
We used the {\sc IRAF} task {\em phot} to perform aperture
photometry on the standard stars
using a sequence of different sized 
apertures of radius 3 to 16 pixels (1\farcs 8-9\farcs 6).
The {\sc IRAF} implementation of the {\sc DAOGROW} algorithm 
(Stetson 1990) was then
used to fit the curve-of-growth function for each image 
and compute the
optimal instrumental magnitudes for the $r=16$ pixel
aperture.  
The resulting list of instrumental magnitudes for the
standard stars was used to define these
transformations from the IRAF instrumental magnitude
system to the standard system:
\begin{eqnarray}
m_U &=& (3.400\pm 0.044) + U + (0.415\pm 0.028)X_U \cr  
    & &    - (0.151\pm 0.077)(U-B)  \cr
    & &    +(0.057\pm 0.053)(U-B)X_U        \cr
m_B &=& (1.425\pm 0.034) + B + (0.236\pm 0.022)X_B \cr
    & &    - (0.023\pm 0.058)(B-V)  \cr
    & &    -(0.049\pm 0.039)(B-V)X_B        \cr
m_V &=& (1.116\pm 0.027) + V + (0.127\pm 0.017)X_V \cr
    & &    + (0.045\pm 0.046)(B-V)  \cr
    & &    -(0.018\pm 0.030)(B-V)X_V        \cr
m_R &=& (1.065\pm 0.018) + R + (0.053\pm 0.011)X_R \cr
    & &    - (0.004\pm 0.052)(V-R)  \cr
    & &    -(0.019\pm 0.034)(V-R)X_R        \cr
m_I &=& (1.630\pm 0.036) + I + (0.014\pm 0.023)X_I \cr
    & &    - (0.176\pm 0.112)(R-I)  \cr
    & &    +(0.079\pm 0.075)(R-I)X_I 
\label{transeq}
\end{eqnarray}
where the lower case $m$'s are the instrumental magnitudes
and where $X$ is the airmass.  
{\em Phot} was also used in a similar manner on the
December 28 KPD 0422+5421
images (the three sequences of $U$, $B$, $V$, $R$, and $I$ exposures).
We prepared these 15 images by using {\sc DAOPHOT} to remove every star
except the image of KPD 0422+5421 and eight bright field stars.  
The transformation  
equations were inverted
to derive the calibrated magnitudes of KPD 0422+5421 and the eight field
stars.  Finally, the mean difference between the 
{\sc DAOMASTER} magnitudes
and the standard $R$ magnitudes for the eight stars was computed and
this offset was applied to every remaining star to arrive at the final
$R$ magnitude light curves.

On December 28.3934 (UT)
we found the following colours and magnitudes for
KPD 0422+5421: $V=14.66 \pm 0.02$, $B-V=0.18\pm 0.02$, $U-B=-0.61\pm
0.04$, $V-R=0.11\pm 0.02$, and $R-I=0.11\pm 0.03$.
For comparison, Downes (1986) measured
$V=14.66$, $B-V=0.20$, and $U-B=-0.65$ (no errors were given).

\section{The Photometric Time Series}

Figure \ref{fig1} shows the light curves of KPD 0422+5421 from each
night (top) and the light curves of two bright field stars. The
standard deviations in these light curves are 0.009 mag for the
brighter comparison star and 0.011 mag for the fainter comparison
star.  The first observations on the fourth night were taken in bright
twilight, and as a result the photometry from that time is rather
noisy.  We therefore excluded the data from the first hour of the
fourth night (our main results are not changed significantly when
these data are included in the analysis).  One can plainly see the
2.16 hour ellipsoidal modulation in the light curves of KPD 0422+5421.
Surprisingly, there is a much longer period signal evident in the
light curves. This signal is especially evident on the third night.
We argue below that this signal is real and not due to an artifact of
the observations or of the data analysis.  Whatever its origin, this
extra signal cannot be neglected since its amplitude is roughly half
that of the ellipsoidal component.  We therefore must remove it before
the ellipsoidal light curve can be properly analyzed.

\subsection{Decomposition of the light curve}

We used an iterative procedure to decompose the light curve into the
ellipsoidal component and the longer-term component.  The
magnitudes were converted to a linear intensity scale and
normalized so that an $R$ magnitude of 14.55 corresponds to an intensity
of 1.0.  There are six  steps in each iteration:
(1) The linearized data were folded on the photometric ephemeris of
KOW and binned into 100 phase bins.  
(2) A W-D model was fitted 
to the folded light curve using a Levenberg-Marquardt
optimization procedure 
(the W-D parameters determined by KOW were used for the initial estimate).
(3) The fitted W-D model
was ``unfolded'' using the KOW ephemeris to yield a curve giving the
intensity as a function of time and subtracted from the
linearised time series data.
(4) The residuals were
searched for periodicities by
computing the variance statistic of Stellingwerf (1978) for trial
periods between 0.01 and 0.5 days.  
(5) The residuals were phased folded on the significant period
found in step (4), binned into 250 phase bins, and 
characterized by a fit using a 2 piece cubic spline.
(6)  The spline fit was unfolded and the resulting curve was
used to detrend the time series data, and the process was started again
at step (1) using the detrended data.

\begin{figure}
\vspace{6cm}
\includegraphics{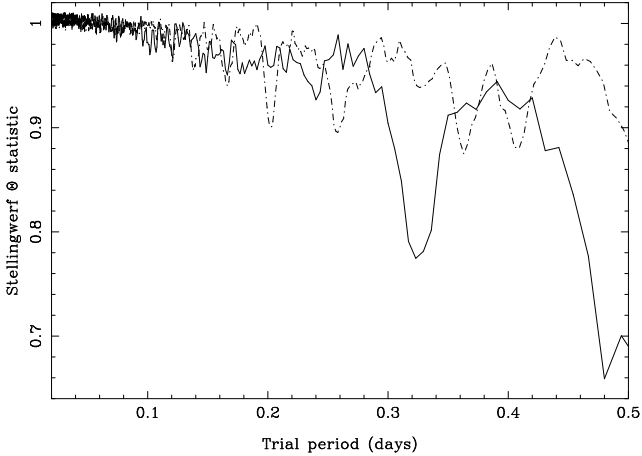}
\caption{The $\Theta$ statistic of Stellingwerf (1978) vs.\
trial period in days for the KPD 0422+5421 light curve with the ellipsoidal
modulation removed (solid line).  
There is a local minimum at a period
of 0.325 days (the minima at 0.245 and 0.481 days are
one cycle per day aliases).  The $\Theta$ statistic for the
comparison star with $R=14.715$ (see Figure 1) is shown as the dash-dotted
line.
}
\label{fig2}
\end{figure}

The results of the period search from step (4) are displayed in
Figure \ref{fig2}.  
There is a local minimum at a trial period of
0.325 days (7.8 hours), meaning there is a coherent modulation at that
period (the local minima at 0.245 and 0.481 days are one cycle per
day aliases of the 0.325 day period).  
The trial period of the local
minimum did not deviate from 0.325 days after the first iteration.
Figure \ref{fig3} 
shows the residuals folded on the 7.8 hour period and
binned into 250 equal phase bins (the phase was adjusted to give the minimum
at phase 0), and the fit using a two piece cubic spline (solid line). 
There is a clear non-sinusoidal modulation with
an amplitude of 1 per cent of the peak flux.  
We found that the character of this 7.8 hour modulation did not noticeably
change after the first iteration.  

The ellipsoidal light curve arrived at
by folding and binning the detrended data did not noticeably change
after the second iteration.
Figure \ref{fig4} 
shows the final ellipsoidal light curve, converted back into
the $R$ magnitude scale, and
the best-fitting W-D model
($i=89^{\circ}$, $M_{\rm WD}/M_{\rm sdB}=1.15$, see below for
details of the fitting procedures).  
The signature of the white dwarf transit across
the face of the sdB star is clearly seen at phase 0.0.  Surprisingly, a
second narrow
dip in the light curve due to the total eclipse of the white dwarf by
the sdB star can be seen at phase 0.5.  The relatively large depth of
this feature ($\approx 0.005$ mag) indicates that the white
dwarf is much brighter than supposed by KOW:
the new W-D fit gives a blackbody temperature of
$T_2= 24,300\pm 2600$~K, compared with the value
of $T_2\approx 4000$~K derived by KOW.  The properties of the white dwarf
are discussed further below.

\subsection{Possible systematic errors}

There is clearly an extra periodic signal in the light curve
of KPD 0422+5421.  Removing this signal gives a cleaner
ellipsoidal light curve.  Is the 7.8 hour modulation intrinsic to 
KPD 0422+5421 or is it an artifact of the observations or reductions?
We consider possible sources of systematic error below.

\begin{figure}
\vspace{6cm}
\includegraphics{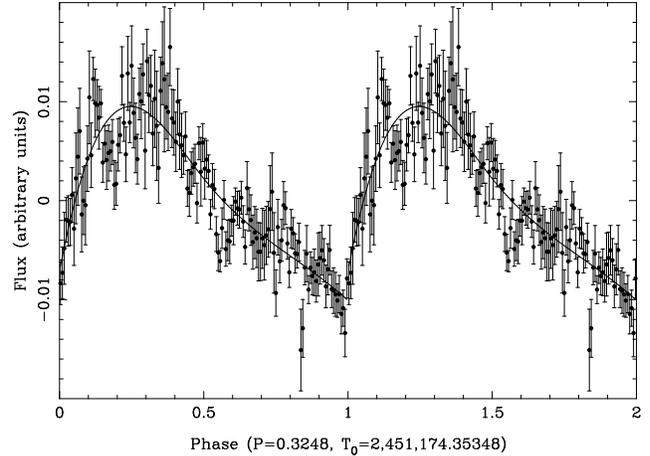}
\caption{The 
residuals are shown folded on a period of 7.8 hours and binned into
150 phase bins, where the points correspond to the median value
within that bin.  The error bars represent the error of the mean
within each bin. 
The solid line is a two piece cubic spline fit. 
Two cycles are shown for clarity.  Note that the $y$-axis is a
linear flux 
scale.}
\label{fig3}
\end{figure}

One could have errors in photometry owing
to changes in seeing and/or changes in
the stellar positions on the CCD (see Frandsen et al.\ 1996).  However, 
since the {\sc DAOPHOT} code uses an
analytic model combined with
an empirical look-up table of residuals when
it fits the PSF to each image, we believe it can robustly
handle changes in the PSF due to guiding errors or changes in seeing.
Furthermore, the 
{\sc ALLSTAR} program recomputes the stellar centroids at each 
iteration based on the most recent PSF.    Thus we do not
expect errors in the
photometry due to seeing changes or shifts in the stellar positions.
Also, one would expect to see spurious signals in the light curves of
other stars if drifts in the stellar positions or changes in seeing
affected the photometry.  However, no spurious signals of comparable amplitude
are evident
in the light curves  of any of the bright field stars.

\begin{figure*}
\vspace{10.7cm}
\includegraphics{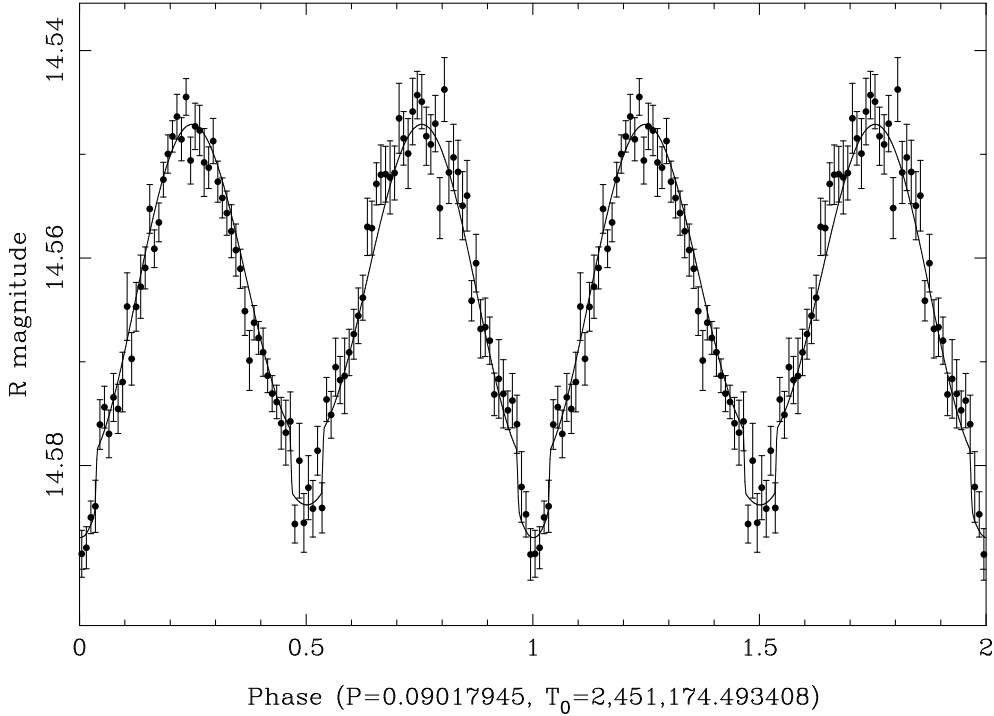}
\caption{The detrended data (i.e.\ the original time series with
the 7.8 hour modulation shown in Figure 
\protect\ref{fig3}
removed)
are shown folded on the orbital period (0.0901795 days) and
binned into 100 bins, where the points correspond to the median value
within that bin.  The error bars represent the error of the mean
within each bin.  The solid line is the best-fitting W-D model (see the 
text).  One can clearly see the signature of 
the transit of the white dwarf
in front of the sdB (phase 0.0)
and the signature of
the occultation of the white dwarf by the sdB
(phase 0.5).
Two cycles are shown for clarity. The $y$-axis here is a magnitude scale.}
\label{fig4}
\end{figure*}

The differential light curves produced by 
the {\sc DAOMASTER} code were made with the assumption that the extinction
(either due to clouds or to changing airmass) is independent of
the colour of the star.  If the extinction did depend on the colour
of the star, then the light curve of KPD 0422+5421 could show a
spurious trend 
since KPD 0422+5421 is bluer than its immediate neighbours ($\approx 0.5$ mag
bluer in $V-R$ than the bluest field stars).
We do not believe our photometry suffers from this problem since
the coefficients of the ``colour-airmass'' terms in
the transformations (equation 1) are all consistent with zero.  Several
of the standard stars we observed are  bluer than KPD 0422+5421
(e.g.\ RU 149 with $B-V=-0.129$ and $U-B=-0.779$, Landolt 1992) and
were observed to fairly high airmass ($\approx 2$), so we would be able
to measure colour-airmass terms that were significantly different from
zero.  As an experiment, we considered only the KPD 0422+5421
observations obtained
at airmass less than 1.4 (1977 images)
and analyzed the residual light curve (i.e.\ the time series data with
the ellipsoidal component removed).  The residuals still showed a periodicity
at 7.8 hours and the residuals folded on 7.8 hours closely
resembled the 
folded curve shown in Figure 
\ref{fig3}.  Thus the 7.8 hour modulation does not
seem to be an artifact of colour-dependent extinction.

None of the other bright field stars show any apparent coherent
periodicity. 
We show  
the $\Theta$ statistic for a comparison star
in Figure 2.  The minimum $\Theta$ value is about 0.89.  Furthermore,
the $\Theta$ values are 
greater than 0.95 at trial periods near
7.8 hours. 
Thus the fact that {\em only} KPD 0422+5421 shows an apparent 
coherent periodicity
is an indication that the cause is not instrumental, and
would have to be related to the observing conditions (for example
the potential systematic error associated with the blue colour 
of KPD 0422+5421 discussed
above).  
However, if the extra signal in the KPD 0422+5421
light curve is indeed an artifact of the observing conditions
and/or procedure, then 
it is not obvious how this signal
could have a periodicity at 7.8 hours.
One might be able to invent 
a periodicity at 7.98 hours (one-third of a sidereal day)
if the signal
were due to the rapid change of airmass as the source sets.  
The fact that the periodicity of the signal is not obviously related to
the rotation of the Earth is further evidence that it is intrinsic to 
KPD 0422+5421 and not an artifact.

\section{Analysis of the Refined Ellipsoidal Light Curve}

\subsection{Model outline}

\begin{figure*}
\vspace{10.7cm}
\includegraphics{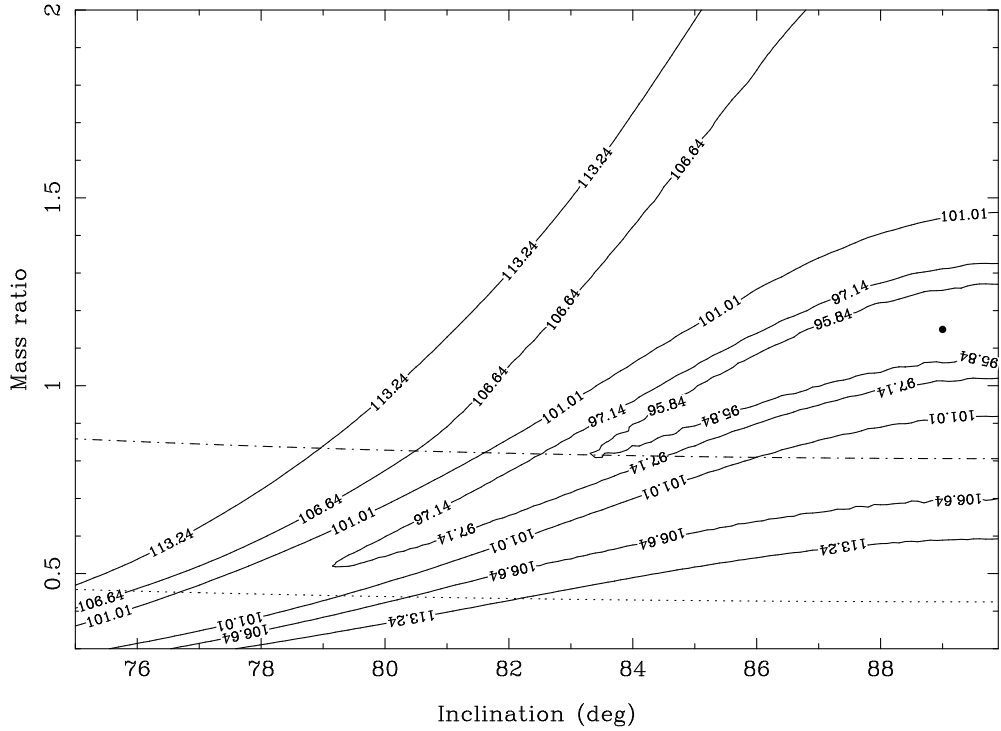}
\caption{A contour map of the $\chi^2$ values is shown (for 100 data bins).  
The filled
circle marks the point with the lowest $\chi^2$ value
($\chi^2_{\rm min}=94.84$ at $i=89.0^{\circ}$ and $Q=1.15$).  The
contour levels shown are for $\chi^2=\chi^2_{\rm min}+1$,
$\chi^2_{\rm min}+2.30$,
$\chi^2_{\rm min}+6.17$,
$\chi^2_{\rm min}+11.8$, and
$\chi^2_{\rm min}+18.4$.  
The latter four contour levels correspond to confidence levels 
of 68.3, 95.4, 99.73 and 99.99 per cent for two parameters of interest.
W-D solutions with mass ratios below the dash-dotted
line have $M_{\rm total}>1.4\,M_{\odot}$, while solutions with mass ratios
below the dotted line have $M_{\rm WD}>1.4\,M_{\odot}$.
}
\label{fig5}
\end{figure*}

With the 7.8 hour modulation  removed from the
time series data, it was possible 
to properly analyze the ellipsoidal light
curve shown in Figure \ref{fig4}.  
We modelled the folded light curve using
``mode 2'' of the W-D code, which is the usual mode used for detached
binaries.  
We assumed a circular orbit and synchronous rotation.  The polar temperature
of the sdB star was fixed at the spectroscopically determined value of
25,000 K (KOW), and we used 
coefficients appropriate for the linear cosine limb darkening law
taken from the tables of Van Hamme (1993).  
The gravity darkening exponents for both stars were set to 1, the usual
values for stars with radiative envelopes (e.g.\ Wilson \& Devinney
1971 and cited references).  [In this particular case, the maximum
temperature change over the star is about 4 per cent.  Although we
do not expect a gravity darkening exponent different than 1, our
fits are not sensitive to small changes 
($\la 10$ per cent) in this parameter.]
Similarly,
the bolometric albedos for reflective heating and re-radiation were also
set to 1.   
(We note that
there is virtually no ``reflection effect'' in KPD 0422+5421
since the sdB star is more than 5.5 mag more luminous than the white dwarf.
Hence our fits are not sensitive to the exact details of the reflection
effect computations.)
The free parameters in the model were the inclination
$i$, the mass ratio $Q\equiv M_{\rm WD}/M_{\rm sdB}$, the surface
potentials $\Omega_{\rm sdB}$ and $\Omega_{\rm WD}$, and the polar
temperature of the white dwarf $T_{\rm pole}$(WD).

The ``grid size integers'' used were $n_1=220$ and $n_2=20$.  A large
value of $n_1$ was needed to obtain a smooth light curve near phase
0.0, when the white dwarf passes in front of the sdB star.
Thus it was necessary
to change an input format statement in the W-D code.  In that same vein,
we also altered the appropriate input format statement to allow
for $\Omega_{\rm WD}$ values greater than 99.9999.

\begin{table*}
\centering
\begin{center}
\caption[]{Parameters for KPD 0422+5421}
\begin{tabular}{lccc} \hline
Parameter &  Value with no  &  Value with white             & Reference \\
          & added constraints &  dwarf mass-radius constraint &        \\
\hline
Orbital Period (days)  &  $0.09017945\pm (1.2\times 10^{-7})$ 
                           & same                            &  This work \\
$T_0$(photo)           &  HJD 2,451,$174.493408\pm 0.000057$ 
                           & same                            & This work \\
$K_{\rm sdB}$ velocity (km s$^{-1})$ & $ 237\pm 18$           
                           & ...                             &     KOW   \\
orbital separation ($R_{\odot}$) & $0.81^{+0.11}_{-0.15}$            
                           & $0.853^{+0.002}_{-0.007}$       &    This work \\
                       &                                     
                       &                                     &       \\
$i$ (degrees)          &  $\ge 84.4$                    
                       &  $\ge 85.6$                       & This work \\
$Q\equiv M_{\rm WD}/M_{\rm sdB}$  &  $1.15^{+0.11}_{-0.34}$          
                       &  $1.07^{+0.09}_{-0.14}$          &  This work \\
$\Omega_{\rm sdB}$     &  $5.75^{+0.27}_{-0.82}$                      
                       &  $5.55^{+0.24}_{-0.30}$    & This work \\
$\Omega_{\rm WD}$     &  $71.73^{+7.50}_{-28.97}$                    
                      &   $62.81^{+9.65}_{-8.58}$           & This work \\
$T_{\rm pole}$(sdB), spectroscopic (K) & $25,000\pm 1500$    
                      &  ...                                 & KOW  \\
$T_{\rm pole}$(WD), W-D model (K) & $22,600\pm 2600$         
                      & same                                & This work  \\
                       &                                     
                      &                                     &       \\
$M_{\rm sdB}$ ($M_{\odot}$)  &  $0.36^{+0.37}_{-0.16}$              
                             & $0.511_{-0.050}^{+0.047}$    &  This work  \\
$M_{\rm WD}$ ($M_{\odot}$)  &  $0.47^{+0.18}_{-0.16}$              
                            &  $0.526^{+0.033}_{-0.030}$    &  This work  \\
total mass ($M_{\odot}$)    &   $0.86^{+0.52}_{-0.35}$      
                            &   $1.026_{-0.036}^{+0.079}$      & This work \\
$R_{\rm sdB}$, polar ($R_{\odot}$)  &  $0.17^{+0.05}_{-0.02}$  
                            &  $0.193^{+0.009}_{-0.008}$    &  This work  \\
$R_{\rm WD}$, polar ($R_{\odot}$)  &  $0.013^{+0.004}_{-0.002}$   
                            &  $0.0145^{+0.0007}_{-0.0007}$    &  This work  \\
$\log g_{\rm sdB}$, W-D model (cgs) & $5.55^{+0.06}_{-0.05}$     
                            & $5.565^{+0.012}_{-0.006}$    &   This work  \\
$\log g_{\rm sdB}$, spectroscopic (cgs) & $5.4\pm 0.1$     
                            &   ...                        &   KOW       \\
\hline
Note:  All errors quoted are 68 per cent confidence.
\end{tabular}
\end{center}
\label{tab1}
\end{table*}

\subsection{Model optimization}

We used three types of optimization schemes to fit the light curves,
a routine adapted
from Bevington (1969) based on the Levenberg-Marquardt
scheme, the ``grid search'' routine adapted from Bevington (1969),
and the ``differential corrections'' program supplied with the
W-D code.  
In our experience a ``brute force'' procedure of searching parameter
space is needed to reliably find the global $\chi^2$ minimum, rather than
just a local minimum.
We began by using several different initial guesses
of parameter values for the Levenberg-Marquardt
optimization routine, including
the parameters found by KOW, and found a set of parameter values
that gave a relatively small value of $\chi^2$.  At this point
it became apparent that the overall light curve is insensitive to the
value of the polar temperature of the white dwarf--changing 
$T_{\rm pole}$(WD) only results in a change of the depth of the
feature near phase 0.5.  We therefore fixed $T_{\rm pole}$(WD)
at its optimal value of 24,300 K, leaving four free parameters.  Next,
we defined a grid of values in the inclination-mass ratio plane
covering a range of $75\le i\le 90^{\circ}$ in steps of $0.1^{\circ}$
and $0.30\le Q\le 2.00$ in steps of 0.1.  We started at the point
in the $i-Q$ plane corresponding to the provisional solution
found above 
($i=82.6^{\circ}$ and $Q=0.76$)
and the routine was run at each point on the grid.  For these
fits, 
the values of $i$ and $Q$ were fixed at the values corresponding to
the grid location, leaving only the two surface potentials as free
parameters.  The grid search routine was then used to optimize the
$\Omega$s.  Figure 5 shows a contour map of the $\chi^2$ values.
We found
the minimum $\chi^2$ value of $\chi^2_{\rm min}=94.84$ for 100
data points occurred at the point $i=89.0^{\circ}$ and $Q=1.15$, relatively
far from our provisional fit found above.  
The standard deviation of the residuals (i.e.\ the data minus the model fit)
is 0.002 mag, which is comparable to the statistical errors of the points
in the binned light curve.

Our values of $i=89.0^{\circ}$ and $Q=1.15$
are significantly different from the values of
$i=78.05\pm 0.50^{\circ}$ and
$Q=0.87\pm 0.15$ found by KOW.   
We believe our present
results are more reliable than those of KOW for two reasons.
First, we have $\approx 36.6$ hours of CCD observations from a 2.1m
telescope
(of which $\approx 8.2$ hours was spent reading out the CCD), compared to
7.9 hours of observations obtained by KOW using a high-speed photometer 
on a 2.1m telescope.  Hence our statistics will be better.
Second, our CCD data are more reliable since we were able to identify
and remove the extra 7.8 hour modulation.  KOW were unaware
of the extra 7.8 hour modulation, and as a result their folded $U$ and
$B$ light curves are likely to be biased and not accurately representative
of the true ellipsoidal light curves.

\subsection{Computation of parameter uncertainties}

\begin{figure}
\centering
\centerline{\epsfxsize=9.1cm
\epsfbox{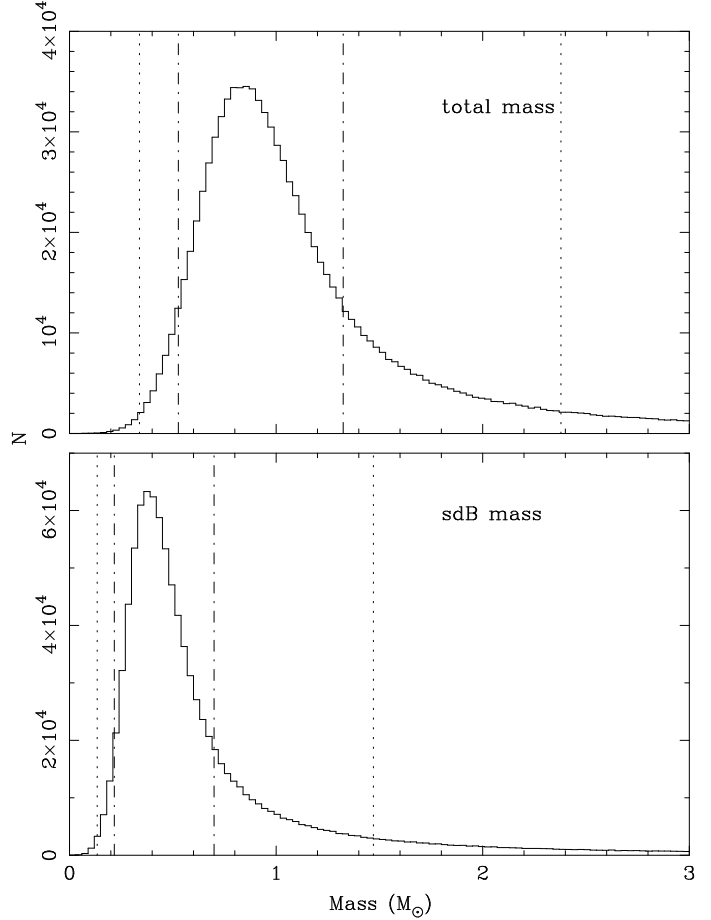}}
\caption{Top:  The marginal distribution of the values for the total
mass as generated using the Monte Carlo procedure.  The dashed-dotted
lines indicate the region that contains 68 per cent
of the area and 
the dotted lines indicate the region that contains 90 per cent
of the area.
Bottom:  Similar to the top, except
the marginal distribution of the values for the sdB mass is displayed.
}
\label{fig6}
\end{figure}

We used a Monte Carlo procedure to compute confidence limits
on the fitted and derived parameters.  At each point in the
$i$-$Q$ plane shown in Figure 5 we have fitted values of
$\Omega_{\rm sdB}$ and $\Omega_{\rm WD}$ and a $\chi^2$ of the fit.
(The fitted values of the $\Omega$s are basically independent of
the inclination and hence are only a function of the mass ration $Q$--larger
values of $Q$ result in larger values for both 
$\Omega_{\rm sdB}$ and $\Omega_{\rm WD}$.)  One can define a region in
the plane corresponding to a certain confidence limit based on the
change in the
$\chi^2$ values, i.e.\ the 68 per cent confidence region is defined by
$\chi^2=\chi^2_{\rm min}+2.34$.  We divided up the plane into 50
regions with region 1 corresponding to confidence limits between
0 and 2 per cent
($\chi^2_{\rm min} \le \chi^2\le\chi^2_{\rm min}+0.041$), 
region 2 corresponding to confidence limits between
2 and 4 per cent
($\chi_{\rm min} +0.041< \chi^2\le\chi^2_{\rm min}+0.081$), 
and so on. The confidence limits defining the last region 
were 98 per cent and 99.99 per cent,
or $\chi_{\rm min}+7.82 \le \chi^2\le\chi^2_{\rm min}+18.43$.
Then 
$2000$ random $i$,$Q$ pairs were selected from each
region and the $\Omega$ values were determined using a 
two-dimensional cubic spline interpolation routine (Press et al.\ 1992).
Finally, for each $i$,$Q$ pair, 10 
values each of the mass function and orbital period
were drawn randomly from appropriate
Gaussian distributions and the
system properties (i.e.\ component masses, orbital separation, etc.)
were computed.  Confidence limits on the parameters were 
computed from the marginal distributions of each parameter.  Figure
6 shows the frequency distributions for the total mass and for the sdB
mass.  The mode of the sdB mass distribution is at $M_{\rm sdB}=0.38
\,M_{\odot}$, 68 per cent
of the values are  in the range
$0.22\,M_{\odot}\le M_{\rm sdB}\le 0.69\,M_{\odot}$
(the lines denoting
the confidence limits intersect the histogram at the same $N$-values), 
and 90 per cent of the values
are in the range
$0.15\,M_{\odot}\le M_{\rm sdB}\le 1.30\,M_{\odot}$.
The corresponding values for the total mass are
$M_{\rm total}=0.84
\,M_{\odot}$ for the mode, 
$0.53\,M_{\odot}\le M_{\rm total}\le 1.34\,M_{\odot}$ for the 68 per cent
confidence interval,
and
$0.39\,M_{\odot}\le M_{\rm total}\le 2.22\,M_{\odot}$
for the 90 per cent
confidence interval.  
Table 1 gives the system parameters for KPD 0422+5421 and their
68 per cent confidence regions.

\subsection{Additional astrophysical constraints}

The 68 per cent
and 90 per cent
confidence intervals on the masses are somewhat large, which
is not surprising considering the relatively large range of mass ratios
allowed by the W-D fits.  There are potentially two separate astrophysical
constraints we can use to limit the parameter ranges, which we describe below.

\subsubsection{The white dwarf mass-radius relation}

It is highly likely that
the unseen star is a white dwarf, and as such has a maximum allowed mass
and a well-defined mass-radius relation.  
However, 
the W-D code makes no assumption as to what the nature of the unseen
star is.  The mass and radius of the unseen star are adjusted as 
necessary when fitting the light curve.  
In our particular situation, 
$\Omega_{\rm WD}$ is positively correlated with $Q$.
Since the mass of the white dwarf is inversely correlated with $Q$
and the radius of the white dwarf is inversely correlated with
$\Omega_{\rm WD}$, we find that the mass of the white dwarf in the W-D
fits
is positively correlated with its radius (see the bottom panel
of Figure 7).
The actual mass-radius relation for white dwarfs is in the opposite
sense:  the radius of a white dwarf get smaller as its mass gets larger.
The form of the mass-radius
relation is model dependent, and we  used the evolutionary tracks of
Althaus \&
Benvenuto (1997, 1998)
and Benvenuto \& Althaus (1999).   These authors have
computed evolutionary
tracks for both CO and He white dwarfs using a range of hydrogen envelope
masses and for a range of metallicities.  For each track, we plot in
the lower panel of Figure 7
the radius of the white dwarf as a function of its mass
when $T_{\rm eff}\approx 35,000$~K and
when $T_{\rm eff}\approx 15,000$~K. Some of the low-mass He models do not
get as hot at 35,000~K.  In these cases we plot the radius 
corresponding to the hottest available temperature.   The
intersection of the theoretical mass-radius band and the W-D
model confidence limits is
relatively small.  The lower envelope of the 
theoretical mass-radius band intersects the 95 per cent
confidence contour at
$M_{\rm WD}=0.42\,M_{\odot}$ and the upper envelope intersects
the same contour at $M_{\rm WD}=0.64\,M_{\odot}$.
W-D fits with 
$M_{\rm WD}\la 0.4\,M_{\odot}$ or with
$M_{\rm WD}\ga 0.7\,M_{\odot}$ can be firmly ruled out since 
the fitted radii are well outside the theoretical mass-radius band.
If we use a ``filter'' in our Monte Carlo simulation and reject
the computed values when the white dwarf mass and radius fall outside
the theoretical mass-radius band, we find the following:
$M_{\rm sdB}=0.511_{-0.050}^{+0.047}\,M_{\odot}$,
$M_{\rm WD} =0.526^{+0.033}_{-0.030}\,M_{\odot}$, 
and
$M_{\rm total}=1.026_{-0.036}^{+0.079}\,M_{\odot}$, 68 per cent
confidence, or 
$M_{\rm sdB}=0.511_{-0.066}^{+0.086}\,M_{\odot}$,
$M_{\rm WD} =0.526_{-0.040}^{+0.056}\,M_{\odot}$,
and
$M_{\rm total}=1.026_{-0.049}^{+0.105}\,M_{\odot}$, 90 per cent
confidence.   The mass of the sdB star is quite consistent
with the ``canonical'' extended horizontal branch mass
of $0.5\,M_{\odot}$ (Caloi 1989; Dorman, Rood, \& O'Connell
1993; Saffer et al.\ 1994).  Furthermore, the total mass of the system
appears to be well below $1.4\,M_{\odot}$.
We list in Table 1 the values of the derived parameters 
(with 68 per cent confidence limits) when the mass-radius
constraint is imposed.

\begin{figure}
\centering
\centerline{\epsfxsize=9.1cm
\epsfbox{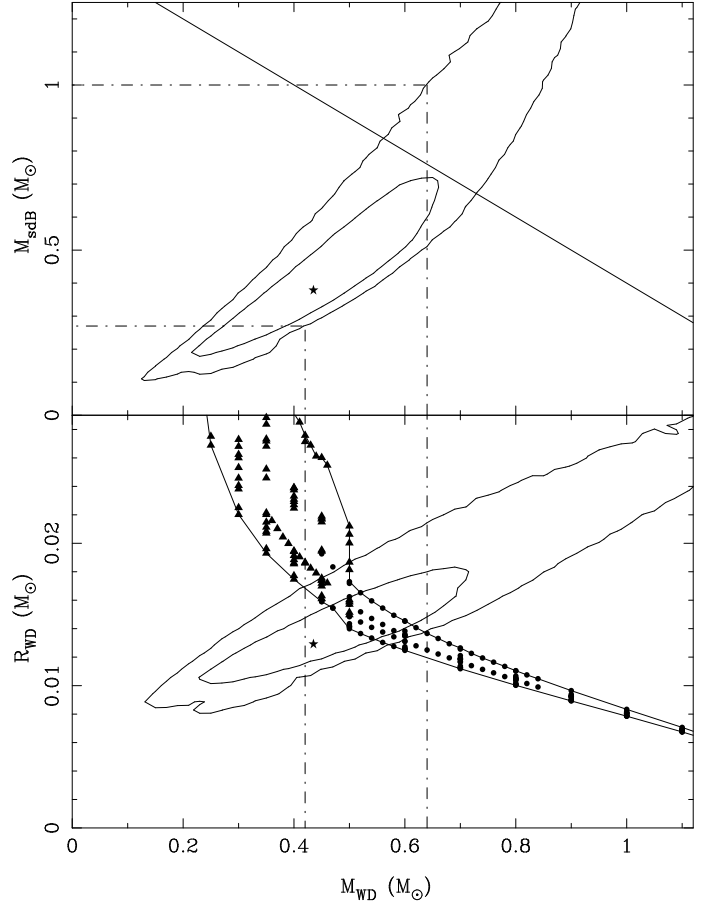}}
\caption{Bottom:  A contour plot of the two-dimensional
frequency distribution of the radius of the white dwarf vs.\ its
mass.  The contour levels indicate the regions that contain 
68 per cent
of the volume and 95 per cent
of the volume.  The star marks the mode.
We also show the theoretical mass-radius relation, taken from the
models of Althaus \& Benvenuto (1997, 1998)
and Benvenuto \& Althaus (1999).  The filled circles
are for CO white dwarfs with $15,000~{\rm K}\le T_{\rm eff}
\le 30,000~{\rm K}$ and with various hydrogen envelope masses
and the filled triangles are for He white dwarfs.  The vertical
dashed-dotted lines indicate roughly where the upper and lower envelopes
intersect the 95 per cent
confidence contour.
Top:  A contour plot of the two-dimensional
frequency distribution of the mass of the sdB star vs.\ the mass of the
white dwarf.  The 68 per cent 
and 95 per cent
confidence regions are displayed.
The diagonal solid line indicates $M_{\rm WD}+M_{\rm sdB}=1.4\,M_{\odot}$.
}
\label{fig7}
\end{figure}

It is quite clear that
imposing the white dwarf mass-radius constraint greatly reduces
the uncertainties on the derived parameters.  
Although the form
of the mass-radius relation is model dependent, the
constraint is fairly robust since the theoretical mass-radius
band is nearly perpendicular to the ridge-line of points derived
from the W-D fits and Monte Carlo simulation.  In addition,
we have used models
with a wide range of hydrogen envelope masses and metallicities, and
we have also used a generous range of temperatures (15,000-35,000~K).

As an aside,
we can use the Althaus \& Benvenuto models to estimate the cooling age
of the white dwarf.
Each Althaus
\& Benvenuto model gives the mass, radius, and temperature of the
white dwarf as a function of the cooling age.
For models where 
$0.50\,M_{\odot}\le M_{\rm WD}\le 0.56\,M_{\odot}$, 
$0.0138\,M_{\odot}\le R_{\rm WD}\le 0.0152\,M_{\odot}$, and
$20,000~{\rm K}\le T_{\rm eff}\le  25,200$~K (i.e.\
the 68 per cent confidence ranges), we find a range of cooling ages 
between
11.3 and  50.1 million years.
The fact that the white dwarf
appears to be relatively young may put constraints on how the present-day
binary formed.  For example, if the white dwarf formed first, then 
the second star would have had
$\la 50$ million years additional time to
evolve into the sdB star, assuming the white dwarf has not somehow
been heated since its formation.

\subsubsection{The radius of the sdB star}

For the second method to potentially reduce the parameter
space found by the W-D fits we can exploit the fact that the mass 
of the sdB star is strongly correlated with its radius 
($\Omega_{\rm sdB}$ is positively correlated with $Q$).  
Hence, the observed rotational velocity of the sdB star
will be correlated with its radius.  One must assume a rotational
period for the sdB star, however.  The usual assumption for
close binary stars is that $P_{\rm rot}=P_{\rm orb}$, i.e.\
synchronous rotation, although this may not necessarily be the
case (see below).  If the rotation is synchronous, then
$V_{\rm rot}\sin i=80$ km s$^{-1}$ when
$M_{\rm sdB}=0.2\,M_{\odot}$ and
$V_{\rm rot}\sin i=125$ km s$^{-1}$ when
$M_{\rm sdB}=0.7\,M_{\odot}$, where the values are computed without
using the white dwarf mass-radius constraint.  Thus a measurement of
$V_{\rm rot}\sin i$ to better than $\approx 10$ km s$^{-1}$
would allow one to reduce the available parameter space for the
mass of the sdB star.   The surface gravity computed from the
W-D fits is also correlated with the computed mass of the sdB star.
We find $\log g=5.46$ (cgs) when 
$M_{\rm sdB}=0.2\,M_{\odot}$ and
$\log g=5.59$ (cgs) when
$M_{\rm sdB}=0.7\,M_{\odot}$, again
where the values are computed without
using the white dwarf mass-radius constraint. 
A spectroscopic measurement of $\log g$ is model dependent and
requires high quality data (one slight advantage of using
$\log g$ to constrain the range of sdB masses is that one
does not need to assume a rotational period for the sdB star).
Unfortunately, 
the spectroscopic measurement given by
KOW of 
$\log g=5.4 \pm 0.1$ does not provide a definitive answer.  The 
$2\sigma$ range covers nearly all of the $\log g$ range 
of the W-D fits and Monte Carlo simulation.

\subsection{A refined orbital ephemeris}

The more precisely defined eclipse profiles also allow us
to determine the orbital phase much more accurately.  The differential
corrections routine of the W-D code can be used to determine the
optimal phase (relative to some assumed time) and its probable error.
The heliocentric
time of the superior conjunction of the sdB star is given
in Table 1.  The formal $1\sigma$ error on this time is 4.9 seconds, 
compared with the error of 43 seconds on that measurement given in
KOW\footnote{Note that the 
error on KOW's value of $T_0$(photo) should be 0.0005 days, not
0.005 days as given.}.  Our determination of the phase combined 
with KOW's
determination leads to the  improved measurement of the
orbital period given in Table 1.   

The accuracy with which we can determine the orbital phase allows
for the  possibility of measuring the change in the binary
period.  A possible cause of a change in the orbital period would be
the loss of orbital angular momentum via
the radiation of gravitational waves.
Ritter (1986)   gives the period derivative due to gravitational
wave radiation as
\begin{equation}
\dot{P}_{\rm GR}=-2.18\times
10^{-14}M_{\rm sdB}M_{\rm WD}(M_{\rm sdB}+M_{\rm WD})^{-1/3}P^{-5/3}
\label{GR}
\end{equation}
where the masses are in solar masses and the period is in days.
Using the masses computed using the white dwarf mass-radius constraint
and $P$ given in Table 1 we find
$\dot{P}_{\rm GR}=-(3.1^{+0.4}_{-0.2})\times 10^{-13}$.  
The phase difference $\Delta\phi_{o-c}$
which accumulates over $N$ orbital cycles
between the observed orbital phase and the expected orbital phase
based on a constant period
is $\Delta\phi_{o-c}\approx 0.5\dot{P}N^2$ (Ritter 1986).
In ten years (40,502 orbital cycles) the accumulated shift will be
$\approx 2$ seconds.
The formal $1\sigma$ error of our present phase determination is
4.9 seconds.   One would need more than twenty years of phase measurements
to begin to detect a marginal phase shift, assuming future measurements
of the phase have similar uncertainties as ours.
The phase might be 
more precisely determined if the ingress and egress portions of the
eclipse profiles were better resolved, although this 
would be a somewhat daunting 
task as the white dwarf takes only $\approx 45$ seconds to move 
a distance equal to its diameter.  Nevertheless, a measurement
of $\dot{P}_{\rm GR}$ would be valuable since it would provide an independent
constraint
on the total mass of the system. This assumes that there is
no other mechanism acting to remove orbital angular momentum.
Note that evolutionary changes of radius of the sdB star on a $10^8$
year time-scale might lead to a $\dot{P}$
term of similar size to
$\dot{P}_{\rm GR}$, if a small portion of the total angular momentum
of the binary is stored in synchronized rotation of the sdB star.

\section{Possible Sources of the 7.8 hour Modulation}

What is the physical cause of the 7.8 hour modulation which we assert is
real?
The white dwarf is too faint to account for the
7.8 hour modulation since the depth of the white dwarf eclipse is
0.006 mag whereas the range of the 7.8 hour modulation is
$\approx 0.02$ mag from minimum to maximum.  Hence the extra 
(non-ellipsoidal) light must be from the sdB star.  We discuss a few
possibilities below.

\subsection{Star spots?}

One  possibility
is that the 7.8 hour modulation is due to a spot on the sdB star
that rotates in and out of view.  In this
case, the rotational velocity of the sdB star would be substantially
slower than the velocity if it were in synchronous rotation:
$V_{\rm rot}\approx 28$ km s$^{-1}$ compared with $V_{\rm rot} = 100$
km s$^{-1}$ for synchronous rotation.  It should be straightforward
to obtain a high resolution spectrum of KPD 0422+5421 and measure
the rotational broadening to see whether the measured rotational velocity
is consistent with synchronous rotation (the resolution of KOW's
spectrum was not sufficient to place any meaningful constraints
on $V_{\rm rot}$).

The W-D fits we have computed all assume synchronous rotation.  The
W-D code allows one to compute models when $P_{\rm rot}\neq
P_{\rm orb}$.
We computed a series of W-D models with the ``$f_1$'' parameter
set to 2.16/7.8=0.277.  The best-fitting model had $i=89.0^{\circ}$,
$Q=1.19$, and $\chi^2=94.83$.  The $\chi^2$ contours in the $i$-$Q$
plane looked similar to the contour plot shown in Figure 5.  Hence our results
are insensitive to changes in the rotational period of the sdB star.
We also experimented with ``spot'' models in the following way:
We added a spot to 
the sdB star and computed several orbital cycles in an attempt
to mimic the time series displayed in Figure 1.   Each model time series
was detrended (using an unspotted model) to isolate the
underlying spot
modulation.  We found that we could not get a  modulation 
with a form matching the shape of the modulation shown
in Figure 3 using a single spot.  Several spots of various
sizes placed at strategic locations on the star may be needed
to produce the ``sawtooth'' pattern that we observe.
Until a definitive measurement of $V_{\rm rot}\sin i$ is made that
demonstrates $P_{\rm rot}\approx 7.8$ hours (i.e.\
$V_{\rm rot}\sin i\approx 28$~km s$^{-1}$ rather than
$\approx 100$~km s$^{-1}$ expected for synchronous rotation), a
more thorough exploration of spot models does not seem warranted
at this time.

\subsection{Pulsation?}

Another possibility is that the modulation could be due to a pulsational
instability in the sdB star.  Recently, a new subclass of sdB stars
known as the EC 14026 stars has been identified
(Kilkenny et al.\ 1997; Koen et al.\ 1997; Stobie et al.\ 1997;
O'Donoghue et al.\ 1997).  The EC 14026 stars show modulations with amplitudes
of $\la 0.01$ mag and periods of $\la 400$ seconds.
In a few cases, these modulations have been identified with low-order
radial and non-radial $p$ and $f$ pulsation modes
(Bill\'eres et al.\ 1997, 1998).      
Many, but not all, of the EC 14026 stars have detectable main sequence
companions (e.g.\ Kilkenny et al.\ 1998 and cited references).  
One system, PG 1336-018, is in a HW Vir type binary with an orbital
period of 0.10 days (Kilkenny et al.\ 1998).  It is not clear 
whether the
other EC 14026 binaries are astrophysically ``close'' (periods less than
a few days) or ``wide'' (separations of greater than a few AU).

KPD 0422+5421, with $T_{\rm eff}\approx 25,000$~K is much
cooler than the EC 14026 stars ($30,000~{\rm K}\la
T_{\rm eff}\la 35,000~{\rm K}$, Billeres et al.\ 1997, 1998).
Although the amplitudes of the
modes observed in the EC 14026 stars are similar to what we observe
in KPD 0422+5421, the periods are much shorter. 
Also, the pulsations observed in the EC 14026 stars have waveforms
that are more sinusoidal, rather than the sawtooth pattern we observed.
Thus if 
KPD 0422+5421 is indeed a pulsator, it probably would be in 
a different class than the EC 14026 stars.

Whatever its cause, the 7.8 hour modulation we observe in KPD 0422+5421
appears to be unique among sdB stars.  
However, we should note an important point.  It is often the
case that in reducing high-speed photometry, long term
trends (e.g.\ timescales of hours) are typically removed
using polynomial prewhitening or other techniques (C. Koen,
private communication).  Thus we encourage observers to
reexamine their existing data and to acquire additional data.  If
the 7.8 hour modulation is a real feature of KPD 0422+5421, then
our observation should be repeatable if given enough care.
If the 7.8 hour modulation is not real, or not strictly periodic,
than that too will
become evident with additional data.

\section{Summary and suggested Future Observations}

We have presented additional and extensive CCD observations of
KPD 0422+5421.  We have discovered that the light curve of this binary
star contains two distinct signals, the 2.16 hour ellipsoidal modulation
and a non-sinusoidal modulation at 7.8 hours.  This extra modulation 
appears to be intrinsic to KPD 0422+5421 and is not an artifact of
the observation or analysis.
We removed this
7.8 hour modulation from the light curve and folded the detrended
curve on the orbital period to obtain the underlying
ellipsoidal light curve.
This folded light curve now clearly shows the signature
of the transit of the white dwarf across the face of the sdB star and
the signature of the total eclipse of the white dwarf (although the latter
feature is somewhat noisier than the former feature).  We modelled the
ellipsoidal light curve with the W-D code and derived  
values of the inclination, the mass ratio, and the $\Omega$ potentials,
and used a Monte Carlo code to compute the uncertainties in the
parameters.  The marginal distributions of the component masses
as computed with the Monte Carlo code have large ``tails'', hence
the parameters have relatively large uncertainties.
If we apply an additional constraint and require the white dwarf's
radius and mass to be consistent with the theoretical mass-radius
relation, we find the derived system parameters have  much smaller
uncertainties.  Using the white dwarf mass-radius constraint we
find that the mass
of the sdB star is somewhat lower than the value determined KOW
and is much
closer to the ``canonical''
mass for extended horizontal branch stars.  The total mass of the
system appears to be less than $1.4\,M_{\odot}$ with high confidence
(99.99 per cent).
We also find that the white dwarf is much hotter (and hence much younger) than
found by
KOW.  The resolved eclipse profiles allow us to
determine the phase accurately, which in turn may allow for the 
measurement of the orbital period derivative after a reasonable amount of 
time.  

Our basic picture of the KPD 0422+5421 binary is secure,
i.e.\ it contains an sdB star with a mass close to $0.5\,M_{\odot}$
paired with a nearly equally massive white dwarf.
Additional spectroscopic
observations should be done to improve the determination
of the velocity amplitude of the sdB star, to determine its rotational
velocity, and to refine the measurements of its effective temperature,
surface gravity, and metallicity
(a good measurement of the He abundance is needed for a precise determination
of $\log g$).  An improved $K$-velocity for the sdB star would
result in a more precisely determined mass function, which in turn would
result in better determined masses.  Accurate values of $V_{\rm rot}
\sin i$ and $\log g$ could be used to reduce the uncertainty in
the mass of the sdB star, independent of the white dwarf mass-radius
relation.   
Further photometric observations should be done in several colours to
verify that the 7.8 hour modulation is coherent and to see if
its character depends on the bandpass, to better sample the ingress and
egress phases of the eclipse profiles, and to provide additional phase
determinations so that the change in the period can be
tracked over time.

\section*{Acknowledgments}

We are pleased to
thank the Kitt Peak director Richard Green for the generous
allocation of
a fourth night of observing time and Doug Williams for his able assistance
at the telescope.  We thank Chris Koen for reading an earlier version 
of this paper and the referee for a detailed and helpful report.

\bsp

\label{lastpage}

\end{document}